\documentclass[10pt]{iopart}
\usepackage{graphicx}
\usepackage{iopams}

\begin{document}

\title[]{AC Magnetometry Loop Tracer Compatible with Magnetic Calorimetry for Power Loss Analysis}

\author{T Veile$^1$, M Harmel$^1$, M Zambach$^1$, P Holm$^1$, F L Durhuus$^1$ and C Frandsen$^1$}

\address{$^1$ Department of Physics, Technical University of Denmark, 2800 Kongens Lyngby, Denmark}
\ead{fraca@fysik.dtu.dk}
\vspace{10pt}

\begin{abstract}
Magnetic nanoparticles (MNPs) have garnered significant attention for various applications in the high-kHz-to-MHz range, although their magnetic characterization at these operational conditions has been limited. However, a number of recent studies have showcased high-frequency and high-field amplitude AC magnetometry loop tracers capable of retrieving the magnetic AC hysteresis curve and the associated magnetic properties. In this paper, we present an easily constructable loop tracer that is retrofitted into an existing AC calorimetry setup. This enables the loop tracer to function simultaneously with the AC calorimetry setup and also to be run as a high-frequency AC susceptometer. The loop tracer is shown to work in the frequency range 160-922 kHz with maximum applied fields from 18 to 46 mT depending on the frequency. An iron oxide nanoflower sample is used to test the loop tracer, showcasing high reproducibility in measured magnetic parameters as well as quantitative agreement between the different measurement methods in the setup.
\end{abstract}

\vspace{2pc}
\ioptwocol

\section{Introduction}
Magnetic nanoparticles (MNPs) are being increasingly studied and developed for a number of different applications including power electronics\cite{yun2014a, zambach2025design}, magnetic hyperthermia for cancer treatment\cite{thiesen2008clinical}, and induction heating of catalytic reactions\cite{vinum2018dual,marbaix2020tuning}. These applications operate the MNPs in a high-frequency applied field in the range of 50 kHz to 1 MHz. Characterizing the MNPs using common quasi-static measurement techniques, such as a vibrating sample magnetometer (VSM) or DC superconducting quantum interference device, is unable to unveil the high-frequency magnetic response of these materials. 

AC calorimetry is a commonly used technique for magnetic hyperthermia studies, where the heat loss of MNPs is characterized by measuring the temperature increase as a function of time in an applied field. This method has been shown to give extremely inconsistent results\cite{wildeboer2014reliable, wells2021challenges}, and it is too insensitive for samples with very low loss, such as those developed for power electronics\cite{zambach2025printable}. AC susceptibility measurements are another common technique, but this method is generally limited to magnetic fields of $<$1 mT, i.e. much lower  than the operational fields in applications such as magnetic hyperthermia and induction-heated catalysis.

This has sparked a surge of studies on an AC magnetometry apparatus called a loop tracer, which is able to measure the hysteresis curve of MNPs at high frequency, up to 1 MHz\cite{garaioMultifrequencyEletromagneticApplicator2014}, and high field amplitudes, up to 160 mT\cite{lenoxHighFrequencyHighFieldHysteresis2018}. The fundamental technique of a loop tracer, using a pick-up coil and compensation coil both inside an AC excitation coil, is not a new idea. The technique was introduced in the early twentieth century for studying rapid changes in magnetic properties\cite{PhysikalischeZeitschrift1907}. In 1996, Slade et al.\cite{sladeCombinationHighsensitivityAlternating1996} showed a combined loop tracer and AC susceptometer working at frequencies up to 2 kHz and 30 mT applied fields. Recently, the research area has grown quickly\cite{connordAircooledLitzWire2014, coissonHysteresisLossesSpecific2017, lenoxHighFrequencyHighFieldHysteresis2018, mattinglyOpensourceDeviceHigh2024, onoderaDynamicHysteresisMeasurement2021} and has contributed to the development of MNPs\cite{boekelheideParticleSizedependentMagnetic2021, cabreraDynamicalMagneticResponse2018, castellanos-rubioOutstandingHeatLoss2019}. 
\begin{figure}%
    \centering
    \includegraphics[width=\linewidth]{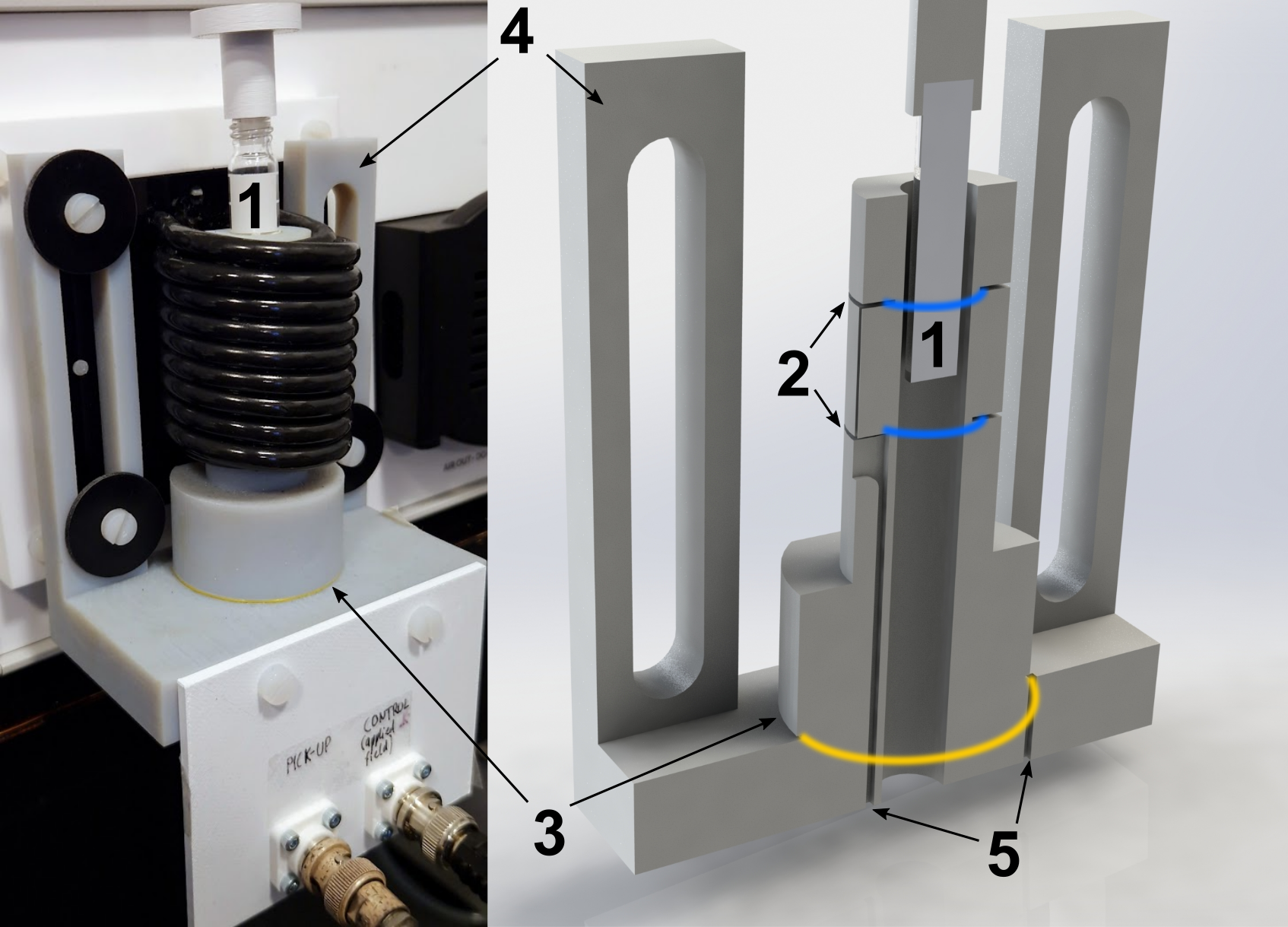}
    \caption{Left: The retrofitted loop tracer (gray) mounted in the excitation coil (black). Right: Cut view of the holder design. The pick-up coil (blue) is centered around the center axis of the excitation coil, only producing a signal when the $B$-field changes in the upper windings. The field coil (yellow) is placed outside the excitation coil to follow the applied field. \textbf{1}: Sample vial, \textbf{2}: Pick-up coil, \textbf{3}: Field coil, \textbf{4}: Mounting rails, \textbf{5}: Wire channels.}
    \label{fig:concept}
\end{figure}
Most notable in relation to this work is the work by Garaio et al., who have built an AC loop tracer, combined it with AC calorimetry\cite{garaioWidefrequencyRangeAC2014, garaioMultifrequencyEletromagneticApplicator2014}, and also implemented external temperature control of the sample to investigate temperature dependent properties of iron oxide nanoparticles\cite{garaioSpecificAbsorptionRate2015, rodrigoi.castellanos-rubioi.garaioe.arriortuao.k.insaustim.oruei.garciaj.a.andplazaolaExploringPotentialDynamic2020}.

In this paper, we present a combined AC loop tracer, AC calorimeter, and AC susceptometer with a broad frequency range from 160 to 922 kHz and maximum applied field amplitudes ranging from 18 to 46 mT. The system enables simultaneous measurement of the AC hysteresis curve and temperature during a running AC calorimetry experiment. This integrated approach allows us to intrinsically capture the temperature dependence of key magnetic properties (susceptibility, remanence, coercivity, and hysteresis area) without the need for external temperature control. By maintaining identical experimental conditions across both techniques, we achieve a direct and reliable comparison between magnetic and calorimetric data. This is a unique approach compared to previous studies, where temperature-dependent AC hysteresis curves were typically obtained by externally adjusting the sample temperature. This continuous application of the AC magnetic field also allows for studying dynamic effects such as chain formation in the sample, potentially influencing the magnetic response on longer timescales, $>$1 second. The device was developed to be retrofitted into a common AC calorimetry setup, the magneTherm, from nanoTherics Ltd.

\section{Principles of Measurements} \label{sec:operation}
The loop tracer consists of a pick-up coil, which has two oppositely wound loops placed inside an excitation coil that generate a signal according to Faraday's law of induction. A sketch of the system is shown in figure \ref{fig:concept}. One loop is wound around the sample, and the other loop is wound around air at the mirrored position in the excitation coil. Without a sample, this results in no signal when applying an AC field, since both coils generate the same magnitude of signal but with opposite signs. Thus, any signal generated from the pick-up coils is due to the magnetization of the sample. Additionally, a coil is placed outside the excitation coil to measure the applied field without any influence from the sample magnetization. This is referred to as the field coil.

The voltages induced from the pick-up coil and field coil can be measured by using an oscilloscope or a lock-in amplifier. In this work, we use a lock-in amplifier, which directly measures the amplitude, phase, and frequency of a signal. The following derivations are based upon measurements with a lock-in amplifier adapted from Barrera et al.~\cite{barreraSpectralAnalysisHysteresis2023}.

\subsection{Signal Analysis}
The applied magnetic field from the excitation coil can be described by a single harmonic oscillation, $H(t)=H_0\cos\left(\omega t+\phi_H\right)$. Due to the non-linear magnetic response, the magnetization of the sample needs to be described by a Fourier series,
\begin{equation}
    M(t)=\sum_{n=1}^N M_n \cos \left( n\omega t +\phi_{M,n} \right).
\label{eq:magnetization}
\end{equation}
The coefficients for the applied magnetic field are determined when measuring the frequency, amplitude, and phase of the first harmonic of the voltage induced in the field coil,
\begin{equation}
    V_H = H^{V} \sin\left( \omega t + \phi^{V}_H \right).
    \label{eq:voltage-field}
\end{equation}
Faraday's law of induction gives the field inducing the voltage,
\begin{equation}
    V_H = - \frac{1}{c_H} \frac{\mathrm{d}H(t)}{\mathrm{d}t} = \frac{H\omega}{c_H}\sin\left(\omega t + \phi_H \right),
    \label{eq:faraday-law}
\end{equation}
where $c_H$ is a geometrical factor that is calibrated (see Section \ref{sec:calibrations}). Comparing the amplitude and phases between equation (\ref{eq:voltage-field}) and (\ref{eq:faraday-law}) gives the expressions,
\begin{equation}
    H = \frac{c_H H^V}{\omega}, \qquad \phi_H = \phi^V_H.
\end{equation}
It is common to use the field signal as reference in the lock-in amplifier, which means $\phi_H^V=\phi_H=0$, but in the following section $\phi_H$ will be included to keep the analysis general.

In a similar manner as with the applied field, all available harmonics are measured for the pick-up coil,
\begin{equation}
    V_M = \sum_{n=1}^N M^{V}_n \cos\left( n\omega t + \phi^{V}_{M,n} \right).
\end{equation}
Utilizing Faraday's law again yields the coefficients for the magnetization,
\begin{equation}
    M_n = \frac{c_M M^V_ n}{n\omega}, \qquad \phi_{M,n} = \phi^V_{M,n}.
\end{equation}

In the following, we derive key magnetic parameters from the measured data.

\subsubsection{Susceptibility} \label{sec:suscept}\hfill\\
Since both amplitude and phase of the signal are measured, it is possible to determine the complex susceptibility. The complex susceptibility is defined as $\chi = \chi'-i\chi''$, where $\chi'$ is the in-phase component , and $\chi''$ is the out-of-phase component of the magnetization in an applied field. This induced magnetization is usually written as,
\begin{eqnarray}
    M(t) = H_0 \sum_{n=1}^N &\chi_n'\cos\left[n(\omega t + \phi_H)\right]  +\nonumber\\ &\chi_n'' \sin\left[n(\omega t + \phi_H) \right].
    \label{eq:magnetization_susept}
\end{eqnarray}
Comparing the coefficients between equations (\ref{eq:magnetization}) and (\ref{eq:magnetization_susept}), makes it possible to determine the susceptibility to any order of harmonic,
\begin{eqnarray}
    \chi'_n =  \frac{M_n}{H_0} \cos\left( \phi_n - n\phi_H \right) \nonumber,\\ 
    \chi''_n = - \frac{M_n}{H_0} \sin\left( \phi_n -n \phi_H \right).
    \label{eq:susceptibility}
\end{eqnarray}
Usually the susceptibility is measured at low fields in the linear regime, where only the first harmonic is non-negligible. 

It is important to note how the lock-in amplifier handles the phase of the reference signal for higher harmonics, as in some cases, the real and imaginary part can crossover\cite{drobacRoleLockinPhase2013}.

\subsubsection{Hysteresis Area}\hfill\\
The area of the hysteresis loop is,
\begin{equation}
    A_{\rm hyst} = \oint M \:\mathrm{d}H.
\end{equation}
Using $\mathrm{d}H = -H_0\omega \sin\left(\omega t + \phi_H \right) \mathrm{d}t$ gives,
\begin{eqnarray}
    A_{\rm hyst} = -\oint  \sum_{n=1}^N &M_n \cos\left(n\omega t + \phi_{M,n} \right)\times\nonumber\\  &H_0\omega \sin \left( \omega t +\phi_H\right) \:\mathrm{d}t. 
\end{eqnarray}
Substituting $x = \omega t$ and using the trigonometric sum identities, $\sin (x+y) = \sin x \cos y + \cos x \sin y$ and $\cos (x+y) = \cos x \cos y - \sin x \sin y$ gives,
\begin{eqnarray}
     A_{\rm hyst} = -\oint  \sum_{n=1}^N M_n H &\left( \cos nx \cos \phi_{n} + \sin nx \sin \phi_{n}  \right) \\ \nonumber
     &\times \left( \sin x \cos \phi_{H} + \cos x \sin \phi_{H}  \right)  \:\mathrm{d}x.
\end{eqnarray}
The cross-terms yield 0, and the sum collapses to $n=1$, since the trigonometric functions form an orthogonal basis. Performing integration for one full cycle, $x=[-\pi, \pi]$ gives the area defined only by the first harmonic,
\begin{equation}
    A_{\rm hyst} = \pi H_0 M_1 \sin\left(\phi_{1} - \phi_H\right).
\end{equation}

\subsubsection{Remanence}\hfill\\
Remanence is the magnetization when the applied field is zero,
\begin{equation}
    H(t) = H \cos (\omega t + \phi_H) = 0.
\end{equation}
This holds true when $\omega t = \pm\pi/2-\phi_H$. Then the magnetic remanence can be calculated from the measured harmonics by,
\begin{equation}
    \pm M_R = \sum_n M_n \cos \left[ n (\pm\pi/2-\phi_H) + \phi_{n}\right] .
    \label{eq:rem}
\end{equation}

\subsubsection{Coercivity}\hfill\\
The coercivity is the field amplitude, where the magnetization is zero,
\begin{eqnarray}
    M(t) &= \sum_n M_n \cos \left[n (\arccos(H_C/H_0)-\phi_H )+\phi_{n} \right]  \nonumber\\  &= 0.
\end{eqnarray}
It is possible to solve this for the first harmonic, $H_C =H_0 \sin(\phi_H-\phi_1)$, but for several harmonics, this equation must be solved numerically.

\section{The Loop Tracer Setup}
The holder for the loop tracer is 3D-printed to be retrofitted into the excitation coil of the commercial magneTherm system from nanoTherics Ltd., a commonly used system in the field of hyperthermia. A cut-view of the holder design can be seen in figure \ref{fig:concept}. The used magneTherm is able to apply fields at five different frequencies in the range of 160-922 kHz with a field amplitude of up to 18-46 mT, which decreases for higher frequency.

For easier construction, the 3D print has built-in channels for the wires that make up the pickup-coil (\textbf{2}) and the field coil (\textbf{3}). The channels are then guided out from the coil through the bottom of the 3D print (\textbf{5}), where they are soldered to BNC connectors. The wires are 0.193 mm in diameter and made of Constantan (a copper nickel alloy) due to its low thermal variation in resistivity. Its relatively high resistivity also helps reduce heating from eddy currents. The wires are insulated with PTFE, giving them a total diameter of 0.7 mm, which eases the construction.

The excitation coil is 54 mm tall with a 25 mm inner diameter. The pick-up coil has a 15 mm diameter and is symmetrically positioned across the coil center, with 20 mm between its upper and lower windings. The field coil, 42 mm in diameter, is placed 25 mm below the excitation coil. The coils are connected to a Zurich Instruments MFLI 5 MHz lock-in amplifier, enabling the system to function as an AC susceptometer when applying small fields. 

To test the capabilities of the loop tracer, a series of measurements are performed on an iron oxide suspension (NF-12). The sample consists of $\alpha$-$\mathrm{Fe_2O_3}$ particles assembled in $\sim$30 nm clusters called nanoflowers\cite{hugounenq2012iron} suspended in water with a weight concentration of 2.1 mg$_{\rm Fe}$/mL. The vial contains a 1 mL cylindrical volume of 13 mm length and 10 mm diameter. An optical fiber, PRB-G40 from Osensa, is used to measure the temperature of the suspension during the measurements.

\section{Corrections and Calibration} \label{sec:calibrations}
\begin{figure}
    \centering
    \includegraphics[width=\linewidth]{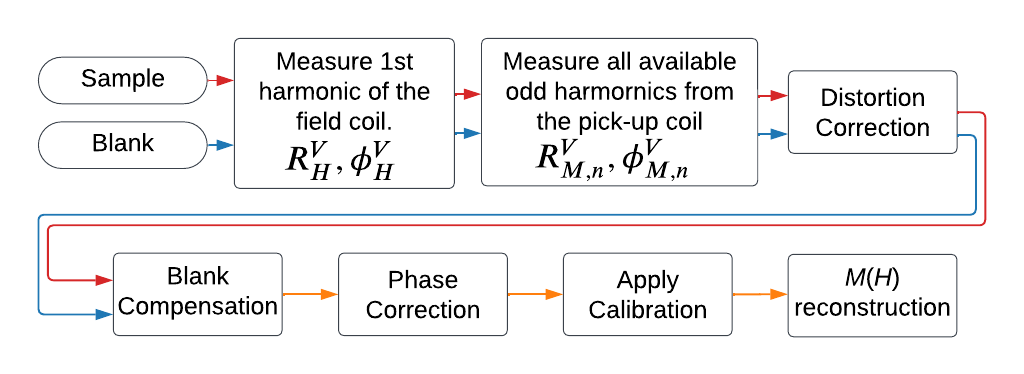}
    \caption{Flow diagram describing the measurement procedure on the loop tracer. Several correction steps have to be implemented after obtaining the raw signal.}
    \label{fig:measurement-procedure}
\end{figure}
The raw signal has to be corrected in several steps to produce reliable results from small variations in the ambient conditions. The measurement procedure can be seen in figure \ref{fig:measurement-procedure} and this section will expand on all the steps leading up to the $M(H)$ reconstruction.
\subsection{Distortion Correction}
\begin{figure}
    \centering
    \includegraphics[width=1\linewidth]{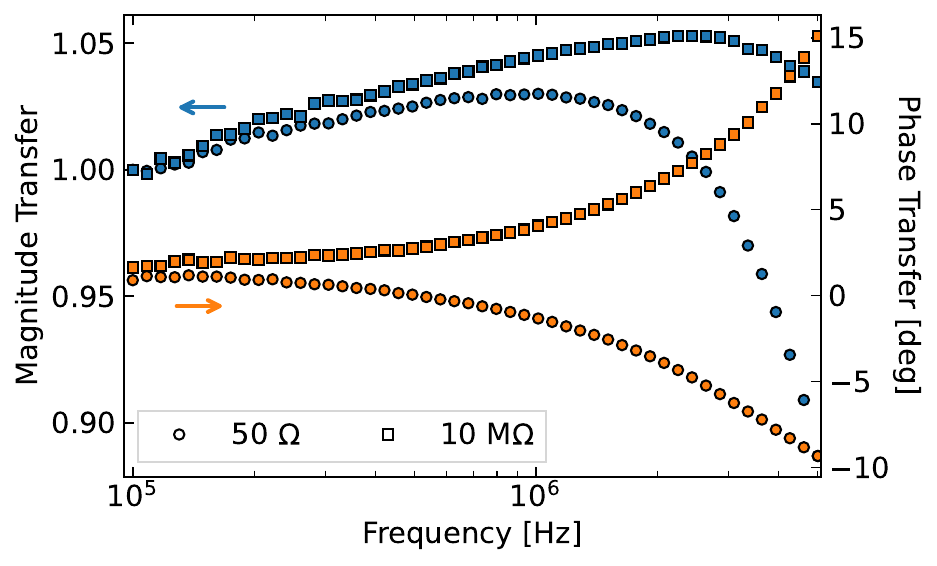}
    \caption{Frequency response of the pick-up coil with two different input impedances. Circles are the magnitude transfer and squares are the phase transfer. The resonant frequency of the pick-up coils lies outside the working range of the loop tracer, minimizing stability issues with the distortion correction.}
    \label{fig:distortion}
\end{figure}

The pick-up coil will distort the signal depending on the frequency due to its complex impedance. This frequency response of the pick-up coil system is determined in a similar fashion to Lenox et al.\cite{lenoxHighFrequencyHighFieldHysteresis2018}, where a signal is generated from a 3-turn coil sweeping through the frequency range of the lock-in amplifier. The known generated signal can then be used to correct any amplitude or phase shift imposed by the pick-up coil. We denote this distortion correction similar to Garaio et al\cite{garaioMultifrequencyEletromagneticApplicator2014}, and it produces two frequency dependent correction factors, magnitude transfer and phase transfer,
\begin{eqnarray}
    &m(f) =& \frac{V_{\rm exp}}{V_{\rm meas}} = \frac{fI_{\rm meas}}{V_{\rm meas}}, \\
    &p(f) =& \phi^{(V)}_{\rm exp} -\phi^{(V)}_{\rm meas} = \phi^{(I)}_{\rm meas} - \pi/2 - \phi^{(V)}_{\rm meas}.
\end{eqnarray}
Figure \ref{fig:distortion} shows the frequency response of our pick-up coil for two different input impedances. The magnitude transfer is normalized to 1 at 100 kHz. Figure \ref{fig:distortion} shows that the pick-up coil system has its resonance frequency outside of the working area for both input impedances. This is beneficial for the stability as small drifts in the distortion will not greatly affect the results.

\subsection{Blank Compensation}
Since the pickup coil is not perfectly symmetric, there will be a small residue signal when no sample is present. Furthermore, the magnetization of the sample holder should be taken into account as both water and glass often exhibit diamagnetic behavior\cite{cini1968temperature}. This system compensates for this by first measuring an empty holder and then subtracting it from the signal of the sample.

It is important to minimize the residue signal in the system, as high residue signals decrease the sensitivity of the equipment. Specifically, when the residue signal is subtracted from the sample signal, the resulting phase becomes more sensitive to minor drifts in the measured phase if the residue signal's amplitude is comparable to that of the sample signal.

The sensitivity limit is assessed by measuring on a blank sample several times to monitor the resulting background signal. The limit is found to be around 1 memu, mostly limited by the uncertainty on the phase.

\subsection{Field, Moment and Phase Calibration}
\begin{figure}
    \centering
    \includegraphics[width=1\linewidth]{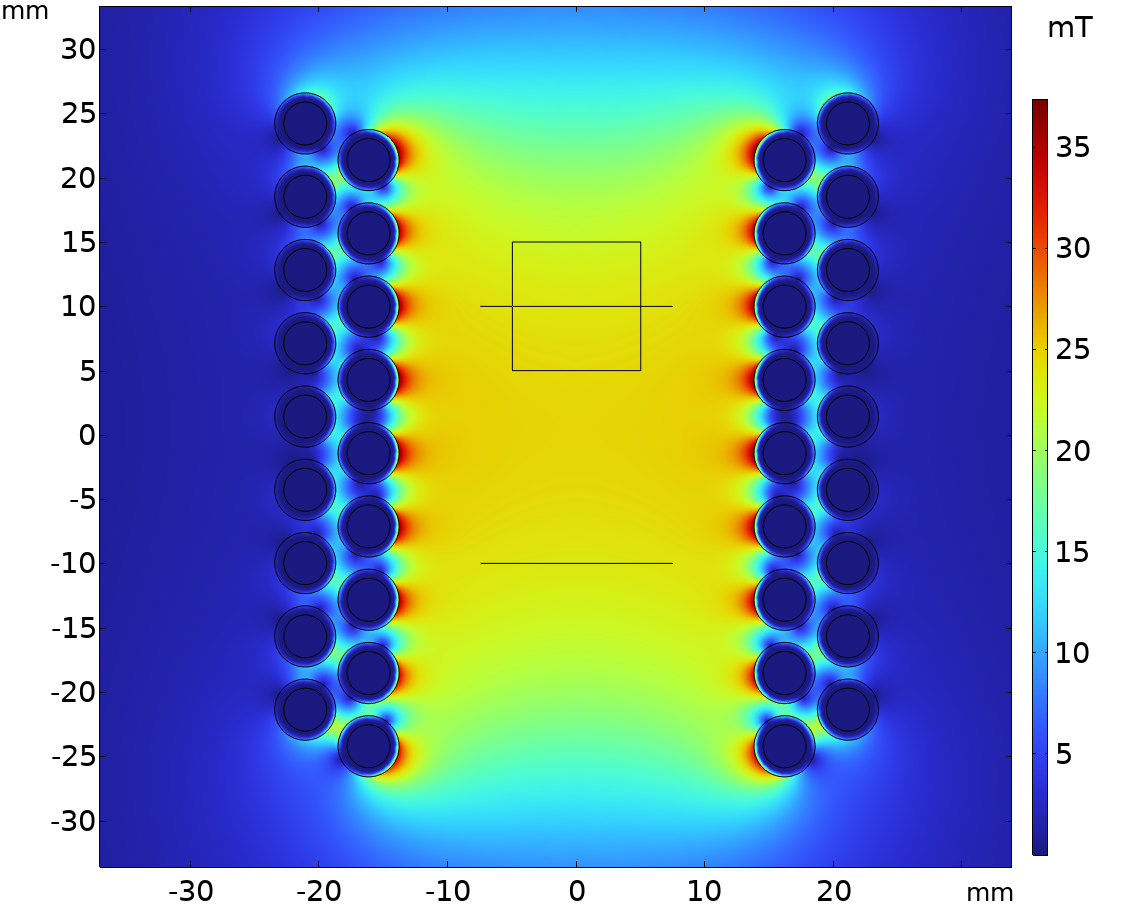}
    \caption{Simulated field strength for the excitation coil in COMSOL. The lines indicate the placement of the pickup coil and the square is the acceptable sample, where the field deviates no more than 5 \%. The sample space is 10 mm high with a diameter of 10 mm.}
    \label{fig:field-simulation}
\end{figure}
The current in the excitation coil was measured with a Rogowski coil at all frequencies, and then the coil was modeled in COMSOL to determine the magnetic field for a given current, which can be seen in figure \ref{fig:field-simulation}. The calibration factor, $c_H$, was found to be consistent across all frequencies and field amplitudes. The value was found to be $c_H = 1.23\times10^{10}$ A (m s V)$^{-1}$ with a standard deviation of 3.13 \%. From the field simulations, we can quantify the field variations across a sample with a 10 mm diameter resulting in 5 \% deviation allowing samples up 10 mm in height. A 10 \% deviation allows for sample heights up to 15 mm. 

The moment and phase are calibrated against a Dy$_2$O$_3$ sample, which is a chemically stable paramagnetic sample with little to no frequency dependency in our working range\cite{tafur2012development}. A paramagnetic sample also has the benefit of a low susceptibility, which makes it possible to neglect demagnetization fields in the measurement. Three calibrations samples were prepared with $\mathrm{Dy_2O_3}$ weights of 2.5823 g, 0.6484 g, and 0.3475 g. The three samples were measured in a vibrating sample magnetometer (VSM) to determine the susceptibility. The calibration factor is found by the ratio between the susceptibility measured by VSM and the first harmonic of the susceptibility measured in the loop tracer, see section \ref{sec:suscept},
\begin{equation}
    c_M = \frac{\chi_{\rm VSM}}{\chi_{\rm LT}^*} = 1.928 \times 10^{-4} \: \mathrm{\frac{A}{m\:s\:V}}.
\end{equation}
$\chi_{\rm LT}^*$ is the uncalibrated susceptibility found with the loop tracer.

The phase of the magnetic response of Dy$_2$O$_3$ compared to the applied field should be zero and we use htis to determine the required phase calibration,
\begin{equation}
    \phi_{\rm cal} = \phi_{M,1} - \phi_H.
\end{equation}
This correction has to be applied for all the harmonics, so the phase becomes,
\begin{equation}
    \phi_{n} = \phi_{n}^* - n\phi_{\rm cal}.
\end{equation}
Here $\phi_{n}^*$ is the uncorrected phase. The phase opening due to the setup is significant, and it is important to correct it. The phase calibration in the system is dependent on the frequency and field, so it is necessary to measure a calibration sample at the same conditions as the sample of interest.

The phase correction is the largest contributor to uncertainty in the loop tracer and will be discussed in further details in section \ref{subsec:drift}.

\section{Results and Discussions}
\begin{figure}
    \centering
    \includegraphics[width=1\linewidth]{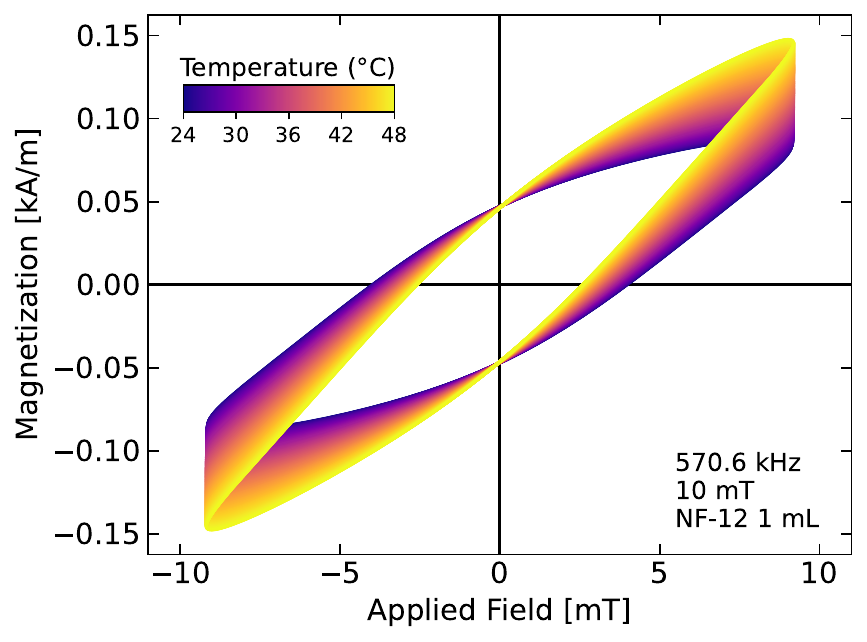}
    \caption{Hysteresis curves for NF-12 measured at different temperatures. The applied field has a frequency of 570.6 kHz and amplitude of 10 mT.}
    \label{fig:hys_nf12}
\end{figure}

Hysteresis curves of the nanoflower sample (NF-12) was measured by the loop tracer at 570.6 kHz and in an applied field of 10 mT. The results are shown in figure \ref{fig:hys_nf12}. It is clear from the measurements that the magnetic response changes significantly with an increase of temperature. The sample appears to soften, as the coercivity decreases while the susceptibility increases. The remanent magnetization remains stable as the temperature increases.

\begin{figure}
    \centering
    \includegraphics[width=1\linewidth]{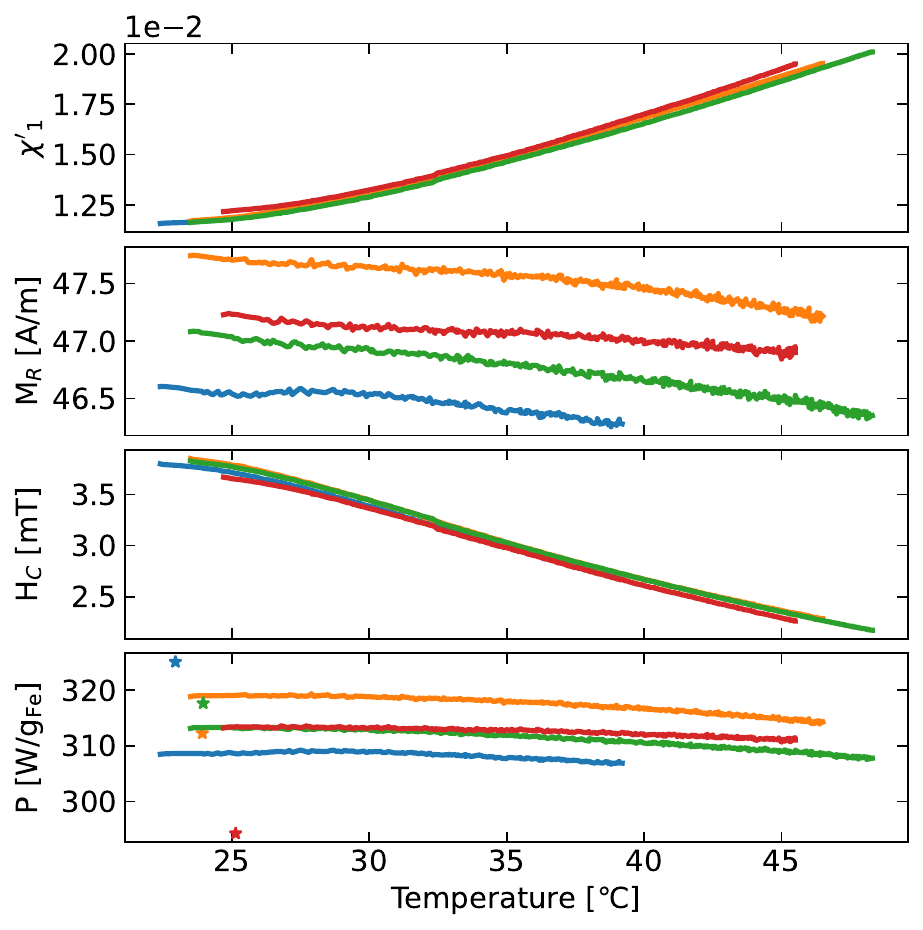}
    \caption{Temperature evolution of all the magnetic properties described in section \ref{sec:operation}: In-phase susceptibility $\chi_1'$, remanence $M_\textrm{R}$, coercivity $H_\textrm{C}$, and hysteresis loss $P$. The colored lines are four repeated loop tracer measurements, each performed at 570.6 kHz and 10 mT, and showing great repeatability (less than 1 \% variation). The stars denote the power loss estimated using AC calorimetry.}
    \label{fig:temp-depend}
\end{figure}

Utilizing the equations derived in section \ref{sec:operation}, we can investigate the change of the magnetic properties as a function of temperature. Figure \ref{fig:temp-depend} shows four repeated measurements at 570.6 kHz and 10 mT over a time period of 8 minutes. The magnetic properties follow the same trend that is apparent in figure \ref{fig:hys_nf12}, but the increase in magnetization keeps the power dissipation rather constant. The power dissipation is determined  simultaneously from AC calorimetry\cite{hanson2023impact} and shown in the figure for comparison. The loop tracer and calorimetry measurements are in good agreement, but the loop tracer displays a higher degree of consistency and gives quantitative information about key magnetic properties and their temperature dependence. 

\subsection{Calibration drift} \label{subsec:drift}
When performing loop tracer measurements simultaneously with AC calorimetry, an issue arises with the reliability for the high frequencies and high field amplitudes. The blank compensation explained in section \ref{sec:calibrations} removes any slight changes to the system as only the difference matters. However, when doing AC calorimetry, the blank measurement is only performed in the beginning. Hence any drifts occurring on the timescale of minutes, will have an effect on the measurements and perceived temperature dependence. The system is tested against this by performing long experiments on the timescale of around 100 seconds on the calibration sample Dy$_2$O$_3$ and monitoring the moment and phase shift for all available field amplitudes and frequencies. 

\begin{figure}
    \centering
    \includegraphics[width=1\linewidth]{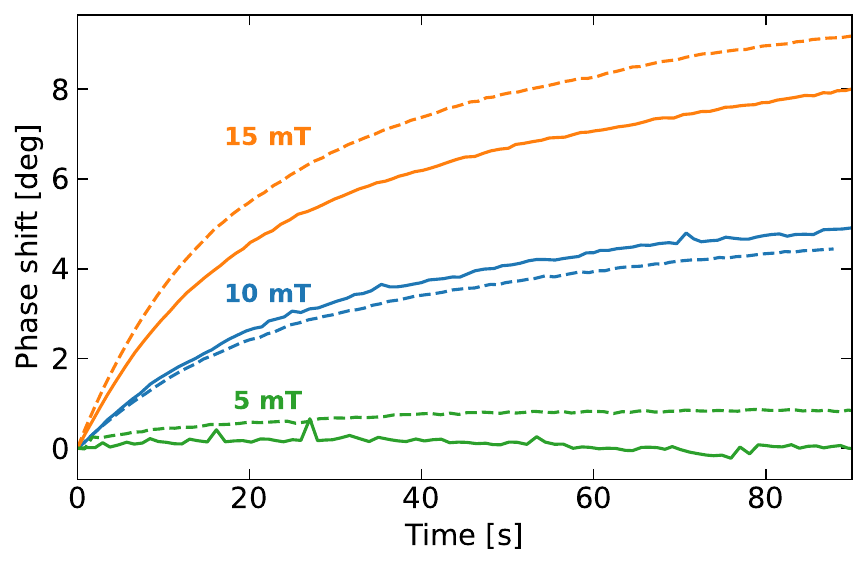}
    \caption{Phase shift calculated from measurements on a Dy$_2$O$_3$ sample as a function of time since blank measurement. The measurements are performed at the highest available frequency of 922.7 kHz. Dashed lines are repeated experiments.}
    \label{fig:phase_drift}
\end{figure}

The moment of the calibration sample is consistent within 5 \% for all frequencies and field amplitudes, but the phase shift is heavily affected at high frequencies. This is shown in figure \ref{fig:phase_drift}, where the phase shift is shown as function of time for different fields at 922.7 kHz. The drift in phase is consistent across repeated measurements, so it is possible to greatly reduce the error by measuring a calibration sample under the same conditions as the sample of interest. Since the drift seems to be heavily frequency dependent, it is probably caused by eddy current heating of the pick-up coils. To mitigate this error, calibration measurements should be performed in the same manner as the sample measurements, allowing the system to reach a comparable temperature state. Redesigning the experiment by measuring several blank measurements during heating would also mitigate these errors, but in turn complicate the AC calorimetry analysis.

\section{Conclusions}
We present a versatile AC magnetometry loop tracer able to measure complete hysteresis curves at high frequencies and field amplitudes. The device can be used simultaneously with AC calorimetry, providing a valuable foundation for comparison of results between the two techniques, while it also intrinsically provides temperature evolution of the magnetic properties. The device is connected to a lock-in amplifier and can run with small field amplitudes as seen in AC susceptometry, thus yielding the complex susceptibility of the sample. In summary, the simultaneous temperature evolution for a range of magnetic properties can be acquired and compared with multiple techniques for validation.

The device is easily constructed by use of a 3D printed holder, and it is retrofittable onto an existing commercially available AC calorimetry setup, which provides a simple upgrade to the numerous existing setups studying power loss using calorimetry.

The biggest obstacle for prolonged experiments is the drift in the phase after blank compensation as the system and pick-up coil heat up. It is possible to mitigate these errors by measuring the calibration sample, Dy$_2$O$_3$, under the same conditions, as the drift is found to be consistent. Designing the experiment with several blank measurements would also greatly reduce this issue.

\ack
We thank Taumoses Legat (grant 2022) and the Independent Research Fund Denmark (grant 0217-00375B E-T-Water) for funding.

\section*{Data Availability Statement}
The data that support the findings of this study are available from the corresponding author upon reasonable request.

\section*{References}
\bibliography{loop-tracer-refs}{}
\bibliographystyle{unsrt}

\end{document}